\documentstyle{article}

\advance\voffset by -2cm

\def\inbar{\vrule height1.5ex width.4pt depth0pt}
\def\rlx{\relax\leavevmode}
\def\I{\leavevmode\hbox{\small1\kern-3.8pt\normalsize1}}
\def\openone{\leavevmode\hbox{\small1\kern-3.3pt\normalsize1}}
\def\Ione{\rlx{\rm 1\kern-2.7pt l}}
\def\Ik{\rlx{\rm I\kern-.18em k}}
\def\IC{\rlx\leavevmode
             \ifmmode\mathchoice
                    {\hbox{\kern.33em\inbar\kern-.3em{\rm C}}}
                    {\hbox{\kern.33em\inbar\kern-.3em{\rm C}}}
                    {\hbox{\kern.28em\sinbar\kern-.25em{\rm C}}}
                    {\hbox{\kern.25em\ssinbar\kern-.22em{\rm C}}}
             \else{\hbox{\kern.3em\inbar\kern-.3em{\rm C}}}\fi}
\def\IP{\rlx{\rm I\kern-.18em P}}
\def\IR{\rlx{\rm I\kern-.18em R}}
\def\IN{\rlx{\rm I\kern-.20em N}}

\def\llsymbol#1{\@llsymbol{\@nameuse{c@#1}}}
\def\@llsymbol#1{\ifcase#1\or {}\or {'}\or {''}\or {'''}\or
   {''''}\or {'''''}\or  \else\@ctrerr\fi\relax}
\newcounter{contador}

\newcommand{\ol}\overline
\newcommand{\ti}\tilde
\newcommand{\wt}\widetilde
\newcommand{\wh}\widehat
\newcommand{\bv}\breve
\newcommand{\dg}\dagger

\newcommand{\be}{\begin{equation}}
\newcommand{\ee}{\end{equation}}
\newcommand{\bl}{\begin{eqnarray}&}
\newcommand{\el}{&\end{eqnarray}}
\newcommand{\bq}{\begin{eqnarray}}
\newcommand{\eq}{\end{eqnarray}}

\begin{document}

%%%%%%%%%%%%%%%%%%%%%%%%%%%%%%%%%%%%%%%%%%%%%%%%%%%%%%%%%%%%%%%%%%
{\hfill \vspace{3mm} }

\begin{center}
{\LARGE {\bf Propagating torsion in 3D-gravity and dynamical mass generation}%
}

\vspace{6mm}

{\large J.L. Boldo$^{{\rm a,}}$}\footnote{%
e-mail address: jboldo@cce.ufes.br, jboldo@cbpf.br}{\large , J.A.
Helay\"{e}l-Neto$^{{\rm b,c,}}$}\footnote{%
e-mail address: helayel@gft.ucp.br}{\large \ and N. Panza$^{{\rm b}}$}

{\it \ \\[0pt]
}$^{{\rm a}}${\it Universidade Federal do Esp\'{\i }rito Santo (UFES)\\[0pt]
Departamento de F\'{\i }sica (CCE)\\[0pt]
Campus Universit\'{a}rio de Goiabeiras - 29060-900 - Vit\'{o}ria - ES-
Brazil.}

\vspace{2mm}$^{{\rm b}}${\it Universidade Cat\'{o}lica de Petr\'{o}polis
(UCP) \\[0pt]
Grupo de F\'{\i }sica Te\'{o}rica \\[0pt]
Rua Bar\~{a}o do Amazonas, 124 - 25685-070 - Petr\'{o}polis - RJ - Brazil.}%
\vspace{2mm}

$^{{\rm c}}${\it Centro Brasileiro de Pequ\'{\i }sas F\'{\i }sicas (CBPF) \\[%
0pt]
Departamento de Teoria deCampos e Part\'{\i }culas (CCP) \\[0pt]
Rua Dr. Xavier Sigaud 150 - 22290-180 - Rio de Janeiro - RJ - Brazil.}

\vspace{2mm}
\end{center}

%\vspace{10mm}

\begin{center}
{\bf Abstract }
\end{center}

In this paper, fermions are minimally coupled to 3D-gravity with dynamical
torsion. A Kalb-Ramond field is also introduced and non-minimally coupled to
the fermions in a gauge-invariant way. We show that a 1-loop mass generation
mechanism takes place for both the 2-form gauge field and the torsion. As
for the fermions, no mass is dynamically generated: at 1-loop, there is only
a fermion mass shift proportional to the Yukawa coupling whenever these
fermions already have a non-vanishing tree-level mass.

%\vspace{10mm}

%%%%%%%%%%%%%%%%%%%%%%%%%%%%%%%%%%%%%%%%%%%%%%%%%%%%%%%%%%%%%%%%%

\section{Introduction}

The mechanism of dynamical (gauge-invariant) mass generation in 2D and 3D
gauge theories has received a great deal of attention over the past two
decades \cite{Schwinger, Pimentel, Delb}. Planar gauge theories, massless or
massive, reveal many peculiarities that justify more detailed investigations 
\cite{Winder}.

Having in mind that topologically massive 3D gravity with dynamical torsion
exhibits a number of peculiar results \cite{Torsion1, Dereli}, we propose
here to pursue the investigation of dynamical mass generation for fermionic
matter and a 2-form gauge potential that interact with gravitational field
whose torsion fluctuations dominate over the metric excitations. In
practice, this means that we adopt 3D space-time as a flat background on
which the torsion degrees of freedom propagate and interact with matter and
gauge fields - it might corresponds to a physical situation where the
torsion field is produced by a cosmological neutrino sea \cite{Letelier}. In
so doing, we actually isolate the genuine contributions of torsion to the
problem of dynamical mass generation.

We initially tackle the issue of mass generation for both the fermion and
the torsion itself. The Yukawa-like interaction between the scalar component
of the torsion - this is a feature of 3D space-time - and the fermion yields
a 1-loop {\it mass correction} for the latter. On the other hand, the
fluctuation of the torsion scalar may acquire a 1-loop mass proportional to
the fermion mass parameter. These results are treated in Section 2.

Proceeding further, in Section 3, we discuss the interaction between a 3D
Kalb-Ramond-type (K-R) field and fermions with non-minimal coupling. As a
free field, the former has no on-shell degrees of freedom, as a peculiarity
of three space-time dimensions. However, the fermions, that couple to both
the torsion and the 2-form gauge potential, induce a 1-loop self-energy
mixing among these fields in the effective action, so that the K-R field
becomes dynamical and acquires a non-trivial mass. We check that unitarity
is ensured at 1-loop and show that the longitudinal degree of freedom
excited at the K-R field does not correspond to a ghost.

Our Concluding Remarks are presented in Section 4.

\section{Matter Coupling}

We start by setting up a few relations and properties of Riemann-Cartan
space-time that we shall make use of throughout this work. The affine
connection is non-symmetric in the first two indices. Its non-symmetric
part, $2\Gamma _{[\mu \nu ]}^{\,\,\,\,\,\,\,\,\,\,\lambda }=T_{\mu \nu
}^{\,\,\,\,\,\,\,\lambda }$, is the torsion tensor \cite{DeSabbata}. In this
space-time, $\Gamma _{\mu \nu }^{\,\,\,\,\,\,\,\lambda }$ can be split as
below: 
\begin{equation}
\Gamma _{\mu \nu }^{\,\,\,\,\,\,\,\lambda }=\left\{ _{\mu \nu }^{\,\,\lambda
}\right\} +K_{\mu \nu }^{\,\,\,\,\,\,\,\lambda },  \label{1}
\end{equation}
where $\left\{ _{\mu \nu }^{\,\,\lambda }\right\} $ is the Christoffel
symbol, completely determined by the metric and its derivatives, and $K_{\mu
\nu }^{\,\,\,\,\,\,\,\lambda }=\frac{1}{2}\left( T_{\mu \nu
}^{\,\,\,\,\,\,\,\lambda }+T_{\,\,\,\mu \nu }^{\lambda }-T_{\nu \,\,\,\mu
}^{\,\,\,\lambda }\right) $ is the contortion tensor, which is antisymmetric
in the last two indices.

This implies that, in such a framework, n-dimensional gravity is described
by two independent objects, namely, the metric and the torsion tensors.

In 3D, the 9 degrees of freedom of the torsion can be covariantly decomposed
into its SO(1,2)-irreducible components: a trace part, $t_{\mu }\equiv
T_{\mu \nu }^{\,\,\,\,\,\,\nu }$, a totally antisymmetric part, $\varphi
\equiv \frac{1}{3!}\varepsilon ^{\mu \nu \lambda }T_{\mu \nu \lambda }$, and
a traceless rank-2 symmetric tensor, $X_{\mu \nu }$. The splitting in the
above components is accomplished by the following relation: 
\begin{equation}
T_{\mu \nu \lambda }=\varphi \varepsilon _{\mu \nu \lambda }+\frac{1}{2}%
(\eta _{\nu \lambda }t_{\mu }-\eta _{\mu \lambda }t_{\nu })+\varepsilon
_{\mu \nu \sigma }X_{\,\,\,\,\,\,\lambda }^{\sigma }.  \label{2}
\end{equation}
The peculiarity of 3D, as compared to 4D-gravity, is the appearance of a
scalar component for the torsion.

In this work, we shall be concerned with the minimal coupling between
fermions and torsion. As we shall soon understand, only the $\varphi $%
-component indeed minimally couples to fermions. On the other hand, it is
worthwhile to mention that the spin-2 component, $X_{\mu \nu }$, plays a
peculiar role when Chern-Simons gravity is considered: it influences the
spin-2 sector of the graviton propagator and it is responsible for the
appearance of a massive ghost-like mode in the spectrum \cite{Torsion1}.

At first sight, the easier way to obtain interactions between torsion and
relativistic fields is through a covariant derivative under the minimal
coupling prescription. As we have seen, the affine connection contains a
torsion-dependent piece, and therefore in each covariant derivative of the
field under consideration there is an interaction with torsion. However, we
must be careful since this procedure, in the case of some fields, may spoil
the gauge invariance of the theory. In fact, an Abelian gauge field, $A_{\mu
}$, cannot interact minimally with torsion while keeping the gauge
invariance requirement (the same occurs with the K-R gauge field, $B_{\mu
\nu }$, which will be object of study in Section 3). For a scalar theory,
the covariant derivative is the usual one, and hence it does not minimally
couple to torsion either. On the other hand, the requirement that the Dirac
equation in a gravitational field preserve local Lorentz invariance yields a
direct interaction between torsion and fermions. In Riemann-Cartan
space-time, the Dirac action for a massive fermion has the form 
\begin{equation}
S_{D}=\int d^{3}x\sqrt{g}\left[ \frac{i}{2}\left( \bar{\psi}\gamma ^{\mu
}D_{\mu }\psi -D_{\mu }\bar{\psi}\gamma ^{\mu }\psi \right) -m\bar{\psi}\psi
\right] ,  \label{3}
\end{equation}
where the covariant derivatives of the spinor fields are given by 
\begin{equation}
D_{\mu }\psi =\partial _{\mu }\psi +\frac{1}{8}B_{\mu }^{\,\,\,ab}\left[
\gamma _{a},\gamma _{b}\right] \psi ,  \label{4}
\end{equation}
\begin{equation}
D_{\mu }\bar{\psi}=\partial _{\mu }\bar{\psi}-\frac{1}{8}B_{\mu }^{\,\,\,ab}%
\bar{\psi}\left[ \gamma _{a},\gamma _{b}\right] ;  \label{5}
\end{equation}
here, Latin indices refer to frame components. $B_{\mu }^{\,\,\,ab}$ are the
components of the spin-connection, 
\begin{equation}
B_{\mu }^{\,\,\,ab}=\omega _{\mu }^{\,\,\,ab}+K_{\mu }^{\,\,\,ab},  \label{6}
\end{equation}
which is the gauge field of the local Lorentz group. $\omega _{\mu
}^{\,\,\,ab}$ is the Riemannian part of the spin-connection: 
\begin{equation}
\omega _{\mu }^{\,\,\,ab}=e_{\mu c}\omega ^{\,\,\,cab}=\frac{1}{2}e_{\mu
c}\left( \Omega ^{cab}+\Omega ^{acb}-\Omega ^{bac}\right) ,  \label{7}
\end{equation}
where $\Omega _{cba}=e_{c}^{\mu }e_{b}^{\nu }\left( \partial _{\mu }e_{\nu
a}-\partial _{\nu }e_{\mu a}\right) $ stands for the rotation coefficients
(see Ref. \cite{DeSabbata} for details); $e_{\mu }^{a}$ stands for the
dreibeins\footnote{%
We are using the following representation for the Dirac matrices in locally
flat 3D space-time: $\gamma ^a=\left( \sigma ^3,\,i\sigma ^1,\,i\sigma
^2\right) ,$ where $\sigma ^i$ are the Pauli matrices.}. Since we are here
mainly interested in the interaction between torsion and fermions, we set $%
g_{\mu \nu }=\eta _{\mu \nu }$. This amounts to saying that we wish to study
torsion effects on a flat background or, rephrasing, the metric fluctuations
are taken to be very weak as compared to the torsion excitations.

After some algebra, the Dirac Lagrangian in the presence of torsion is given
by, 
\begin{equation}
{\cal L}_{0}=\bar{\psi}\left( i\gamma ^{\mu }\partial _{\mu }\psi -m\right)
\psi +\frac{i}{4}K_{\mu \nu \lambda }\bar{\psi}\gamma ^{\left[ \mu \right.
}\gamma ^{\nu }\gamma ^{\left. \lambda \right] }\psi ,  \label{8}
\end{equation}
where $\gamma ^{\left[ \mu \right. }\gamma ^{\nu }\gamma ^{\left. \lambda
\right] }$ means the totally antisymmetric product of three Dirac matrices.
Using the identity, $\varepsilon _{\mu \nu \lambda }\gamma ^{\left[ \mu
\right. }\gamma ^{\nu }\gamma ^{\left. \lambda \right] }=-i3!{\bf 1}$, the
interaction term can be shown to simplify to: 
\begin{equation}
\frac{3}{4}\varphi \bar{\psi}\psi .  \label{9}
\end{equation}
Thus, the interaction of torsion and fermions is seen to be a Yukawa
coupling. We would like to point out that a detailed discussion on the
coupling of Dirac fermions to gravity is presented in the works of Ref. \cite
{Mielke}. In our case, since we are confined to 3D, chiral fermions cannot
appear and our discussion on the dynamical mass generation does not involve
the consideration of the anomaly problem.

Propagating and self-interactions terms for torsion, if not introduced by
hand, according to the arguments of Refs. \cite{Caroll, Shapiro}, may be
justified on more geometrical grounds, as coming from powers of the Ricci
tensor and curvature scalar, as it can be readily seen with the help of:

\begin{eqnarray}
R_{\mu \nu } &=&\frac{1}{2}\left( \varepsilon _{\mu \nu
}^{\,\,\,\,\,\,\lambda }\partial _{\lambda }\varphi -\eta _{\mu \nu }\varphi
^{2}\right) ,  \label{10} \\
R &=&-\frac{3}{2}\varphi ^{2},  \label{11}
\end{eqnarray}
then 
\begin{equation}
R^{\mu \nu }R_{\mu \nu }=\frac{1}{2}\partial _{\mu }\varphi \partial ^{\mu
}\varphi +\frac{3}{4}\varphi ^{4}.  \label{12}
\end{equation}
where we set $g_{\mu \nu }=\eta _{\mu \nu }$ (flat metric background); the
metric fluctuations are neglected as compared to scalar torsion excitations,
since we are mainly concerned with the interaction between torsion and
fermions. Therefore, we first consider the following action 
\begin{equation}
S=S_{D}+S_{R},  \label{13}
\end{equation}
where $S_{D}$ is given by (\ref{3}) and $S_{R}$ is 
\begin{equation}
S_{R}=\int d^{3}x\sqrt{g}\left[ aR+b\left( R^{\mu \nu }R_{\mu \nu }-\frac{1}{%
3}R^{2}\right) \right] .  \label{14}
\end{equation}
Finally, we can write , in a flat background, the Lagrangian for fermions
and torsion as follows: 
\begin{equation}
{\cal L}=\frac{1}{2}\left( \partial _{\mu }\phi \partial ^{\mu }\phi -\mu
^{2}\phi ^{2}\right) +\bar{\psi}\left( i\gamma ^{\mu }\partial _{\mu }\psi
-m\right) \psi +\alpha \phi \bar{\psi}\psi ,  \label{15}
\end{equation}
where we have redefined the field $\varphi \rightarrow \phi =\sqrt{b}\varphi 
$, so that $\mu ^{2}=\frac{a}{b}$ and $\alpha =\frac{3}{4\sqrt{b}}$. It is
worthy to point out that the free action for the scalar is nothing but the
dimensionally reduced form of the torsion action proposed in the Ref. \cite
{Shapiro}. Notice that we must require that $b>0$ and $a>0$ to ensure that $%
\mu ^{2}$ be positive and that $\alpha $ be real. Moreover, the canonical
dimensions of the fields and parameters are given below: 
\begin{equation}
\left[ \phi \right] =\frac{1}{2};\;\;\left[ \psi \right] =1;\;\;\left[
b\right] =-1;\;\;\left[ a\right] =1.  \label{16}
\end{equation}

In the work by Baekler and Mielke \cite{Baekler}, one discusses the most
general gauge-invariant action for gravity with torsion. In our case, since
we have in mind to keep only torsion excitations (we neglect the vielbein
fluctuations), we do not need to add up the translational Chern-Simons term
discussed in such work, since the one would amount to a linear term in the
torsion.

The particular combination of the curvature square terms in the action (\ref
{14}) suppresses the quartic scalar self-interaction term. Also, we notice
that, for massless fermions ($m=0$), the Lagrangian (\ref{15}) has parity ($%
{\cal P}$) symmetry, since $\phi $ (pseudo-scalar) and $\bar{\psi}\psi $ are
not separately invariant under this symmetry operation in 3D, namely 
\begin{eqnarray}
&&\,\,\,\,\,\,\,\phi \stackrel{\cal P}{\rightarrow }-\phi ,  \label{17} \\
&&\bar{\psi}\psi \stackrel{\cal P}{\rightarrow }-\bar{\psi}\psi ,  \nonumber
\end{eqnarray}
so that $\phi \bar{\psi}\psi \stackrel{\cal P}{\rightarrow }\phi \bar{\psi}%
\psi $. We shall come back to this point later on, in connection with the
question of fermion mass generation.

Now, we can extract the Feynman rules for the theory. The propagators for
the scalar and spinor fields that stem from the free action are as follows: 
\begin{equation}
i\Delta _{0}\left( p\right) =\frac{i}{p^{2}-\mu ^{2}}\;\;\;\;{\rm {and\ \ \ }%
}\ iS_{0}\left( p\right) =\frac{i}{\not{p}-m}.  \label{18}
\end{equation}
Furthermore, the fermion-torsion vertex is given by 
\begin{equation}
V\left( \psi ,\phi \right) =i\alpha .  \label{19}
\end{equation}
Since $\alpha $ carries dimension $\frac{1}{2}$, one can see that this
vertex is super-renormalisable. In fact, a power-counting analysis shows
that the superficial degree of divergence of primitively divergent graphs, $%
\delta $, decreases as the number of vertices increases: 
\begin{equation}
\delta =3-\frac{1}{2}V_{n}-\frac{1}{2}E_{\phi }-E_{\psi },  \label{20}
\end{equation}
where $V_{n}$ is the number of vertices, while $E_{\phi }$ and $E_{\psi }$
are the external lines associated to $\phi $ and $\psi $, respectively.

From these results, we turn into the calculation of the self-energy
corrections for the bare propagators and discuss the mechanism of mass
generation for massless fields, by looking at the poles of the 1-loop
corrected propagators. For the fermion self-energy graph, we find: 
\begin{equation}
-i\Sigma \left( p\right) =\left( i\alpha \right) ^{2}\left( i\right)
^{2}\int \frac{d^{3}l}{\left( 2\pi \right) ^{3}}\frac{\left( \not{p}-\not%
{l}+m\right) }{\left[ \left( p-l\right) ^{2}-m^{2}\right] \left( l^{2}-\mu
^{2}\right) }.  \label{21}
\end{equation}
As we are interested in reading off a mass generation, we are allowed to set
the external momentum to zero. In so doing, we find 
\begin{equation}
-i\Sigma \left( 0\right) =i\frac{\alpha ^{2}m}{4\pi \left( \left| m\right|
+\left| \mu \right| \right) }.  \label{22}
\end{equation}
The insertion of this 1-loop graph leads to the following 1-loop corrected
propagator: 
\begin{eqnarray}
iS &=&iS_{0}+iS_{0}\left( -i\Sigma \right) iS_{0}+...  \label{23} \\
&=&\frac{i}{\not{p}-m-\Sigma }.  \nonumber
\end{eqnarray}
The above result shows that, if we had started with a massless torsion%
\footnote{%
Torsion with vanishing mass yields the same infra-red problem as found in QED%
$_3.$}, the fermion self-energy correction (\ref{22}) would still be
different from zero and given exclusively in terms of the Yukawa coupling, $%
\alpha $. However, we warn that this result should not be interpreted as a
dynamical mass generation for the fermion, since taking the limit $%
m\rightarrow 0$ after perturbative calculations have been done is not a
consistent procedure \cite{Cima}; had we started off with massless fermions
and torsion ($m=\mu =0)$, we would have found 
\begin{equation}
-i\Sigma \left( p\right) =\frac{\alpha ^{2}}{16}\frac{\not{p}}{\sqrt{p^{2}}},
\label{24}
\end{equation}
and so no mass would be dynamically generated for the fermion, what would be
consistent with the parity invariance of models with massless fermions in
3D: no parity-anomaly should be induced perturbatively in a model with
massless fermions. Therefore, from (\ref{22}) and $\mu =0$, one obtains a
modified propagator (\ref{23}) which has a massive pole at $p^{2}=\left( m-%
\frac{\alpha ^{2}}{4\pi }\right) ^{2}$.

Also, the torsion self-energy is calculated, and we get 
\begin{eqnarray}
i\Pi _{\phi }\left( p\right) &=&-\left( i\alpha \right) ^{2}\left( i\right)
^{2}\int \frac{d^{3}l}{\left( 2\pi \right) ^{3}}\frac{tr\left[ \left( \not%
{l}+m\right) \left( \not{l}-\not{p}+m\right) \right] }{\left(
l^{2}-m^{2}\right) \left[ \left( l-p\right) ^{2}-m^{2}\right] },  \label{25}
\\
i\Pi _{\phi }\left( 0\right) &=&-i\frac{m\alpha ^{2}}{\pi }.  \nonumber
\end{eqnarray}

Similarly, the modified torsion propagator is obtained by summing up the
series 
\begin{eqnarray}
i\Delta &=&i\Delta _0+i\Delta _0\left( i\Pi _\phi \right) i\Delta _0+...
\label{26} \\
&=&\frac i{p^2-\mu ^2+\Pi _\phi }.  \nonumber
\end{eqnarray}

Here, we have another mass correction, at this time for the pole of the
torsion propagator. From this result, we conclude that, whenever the torsion
field mediates the interaction between massive fermions, it exhibits a short
range behavior.. In other words, if we consider, from the beginning, a
massless torsion theory by setting $a=0$, its 1-loop corrected propagator
exhibits a pole at $p^{2}=\frac{m\alpha ^{2}}{\pi }$. We should stress that
dynamical mass generation for torsion requires its coupling to {\it massive
fermions}.

To conclude this section, we would like to make a short comment on the
possibility of getting mass for the fermion by means of parity spontaneous
breaking in our model. With massless fermions, having in mind that $\phi ^4$%
- and $\phi ^6$- terms could have been introduced via $R^{\mu \nu }R_{\mu
\nu }$, $R^2$ and $R^3$ the most general Higgs-like potential in 3D comes
out (6-th order in $\phi $). Setting suitably the parameters such that the $%
\phi $-field acquires a non-vanishing vacuum expectation value (v.e.v),
fermions would become massive upon the spontaneous breaking of ${\cal P}$ by
means of such a v.e.v. \cite{Torsion3}. This shows that the fermions may
acquire mass already at tree-level and, according to our previous
discussion, 1-loop effects will simply provide a shift in the pole of the
fermion propagator.

So far, we have been dealing with a theory involving torsion and a spinor
field. Next, we wish to include the K-R field and to study a possible way to
endow it with a mass and a dynamical character by means of the
fermion-mediated interaction that couples it to torsion.

\section{Mass Generation for the K-R Field}

Following the same procedure as in the previous section, we shall now
examine the behavior of the antisymmetric K-R field, $B_{\mu \nu }$, in 3D
and the mechanism of mass generation for such a field whenever fermions are
considered. The motivation to study 2-form gauge fields in 3D may be
justified by the interest of studying supersymmetric self-dual vortices with
anomalous magnetic moment coupling \cite{Alvaro} and supersymmetric cosmic
string configurations, as recently discussed in the work of Ref. \cite{Cris}.

We start from the well-known free action for this field:

\begin{equation}
S_{G}=\frac{1}{6}\int d^{3}x\sqrt{g}G_{\mu \nu \lambda }G^{\mu \nu \lambda },
\label{27}
\end{equation}
where $G_{\mu \nu \lambda }$ is the field-strength, written in terms of $%
B_{\mu \nu }$ as\footnote{%
The action is invariant under the local gauge transformation $\delta B_{\mu
\nu }\left( x\right) =\partial _\mu \zeta _\nu \left( x\right) -\partial
_\nu \zeta _\mu \left( x\right) $, where $\zeta _\mu $ is an arbitrary
vector.} 
\begin{equation}
G_{\mu \nu \lambda }=\partial _{\mu }B_{\nu \lambda }+\partial _{\lambda
}B_{\mu \nu }+\partial _{\mu }B_{\lambda \nu }.  \label{28}
\end{equation}

In order to obtain the free propagator of the theory, one must to fix this
gauge invariance; we choose the following gauge-fixing term 
\begin{equation}
S_{gf}=\beta \int d^{3}x\sqrt{g}\left( \eta ^{\mu \nu \lambda }\tilde{\nabla}%
_{\nu }B_{\lambda }\right) ^{2},  \label{30}
\end{equation}
where $\eta ^{\mu \nu \lambda }=\frac{\varepsilon ^{\mu \nu \lambda }}{\sqrt{%
g}}$ and $B_{\mu }=\frac{1}{3!}\varepsilon _{\mu \nu \lambda }B^{\nu \lambda
}$ is the dual of $B_{\mu \nu }$. The tilde on $\nabla _{\mu }$ means that
we are considering only the Riemannian part of the connection in the
covariant derivative. This gauge-fixing dictates a ghost term in the action.
However, since we have $\tilde{\nabla}_{\mu }$, rather than $\nabla _{\mu }$%
, the ghosts do not couple to torsion, but exclusively to the metric degrees
of freedom of the gravitational sector. Nevertheless, since we are assuming
that the latter are not excited, the ghost-gravity coupling introduced by
eq. (\ref{30}) need not be taken into account.

It is worthy to remind that, in 4D, $B_{\mu \nu }$ has one on-shell degree
of freedom and its main application is in connection with the gauge
invariant mechanism of mass generation for the electromagnetic field $A_\mu $
through the topological coupling term $\varepsilon ^{\mu \nu \kappa \lambda
}F_{\mu \nu }B_{\kappa \lambda }$. However, in 3D, the free $B_{\mu \nu }$%
-field does not have on-shell degrees of freedom; therefore, it appears only
in the internal lines of Feynman diagrams describing physical amplitudes.

Then, following our study of dynamical mass generation in 3D theories, we
propose (based on the work of Re. \cite{Alvaro}) the following interaction
term between fermions and $B_{\mu \nu }$, without spoiling gauge invariance: 
\begin{equation}
S_{I}=\xi \int d^{3}x\varepsilon ^{\mu \nu \lambda }\partial _{\mu }B_{\nu
\lambda }\bar{\psi}\psi ,  \label{31}
\end{equation}
in an attempt to obtain a (dynamical) on-shell degree of freedom for $B_{\mu
\nu }$. Actually, the interacting term in (\ref{31}) naturally shows up in
an N=2 supersymmetric model where a non-minimal magnetic moment coupling
favors the formation of topological vortices \cite{Alvaro}.

Perhaps, it would be advisable, for the sake of clarity, to emphasize that $%
B_{\mu \nu }$ can couple only non-minimally to fermions, since the latter do
not transform under the $U\left( 1\right) \,$- symmetry associated to $%
B_{\mu \nu }$. So, the only way to have a gauge-invariant coupling is
through its field strength, $G_{\mu \nu \lambda }$. On the other hand, as
long as torsion is concerned, its non-minimal coupling to fermions yield
dynamical mass generation for the latter \cite{Torsion3}; however, they do
not contribute 1-loop corrections to the mass of the K-R field - this is why
we do not take such a non-minimal coupling into account here.

Here $\xi $ carries dimension $-\frac{1}{2}$, and hence the
renormalizability of the theory is certainly lost. So, one can consider this
only as a low-energy effective theory. In this sense, we begin with a
Lagrangian in the flat space-time background, which becomes from the sum of (%
\ref{13}), (\ref{27}), (\ref{30}) and (\ref{31}), namely 
\begin{equation}
{\cal L}=\frac{1}{2}\left( \partial _{\mu }\phi \partial ^{\mu }\phi -\mu
^{2}\phi ^{2}\right) +\left( \partial _{\mu }B^{\mu }\right) ^{2}+\beta
\left( \varepsilon ^{\mu \nu \lambda }\partial _{\nu }B_{\lambda }\right)
^{2}+\bar{\psi}\left( i\gamma ^{\mu }\partial _{\mu }-m\right) \psi +\alpha
\phi \bar{\psi}\psi +\xi \partial _{\mu }B^{\mu }\bar{\psi}\psi ,  \label{32}
\end{equation}
where the bilinear sector of the Lagrangian leads, besides the propagators
listed in (\ref{18}) for torsion and fermion fields, to the following $%
B_{\mu }$-propagator 
\begin{equation}
iG_{0}^{\mu \nu }\left( p\right) =\frac{i}{2p^{2}}\left( \frac{1}{\beta }%
\theta ^{\mu \nu }+\omega ^{\mu \nu }\right) ,  \label{33}
\end{equation}
$\theta _{\mu \nu }$ and $\omega _{\mu \nu }$ being the transverse and
longitudinal projectors in the space of vectors, respectively. Notice that
the K-R field appears in (\ref{32}) through its dual ($B_{\mu }$).

Now, we can verify the absence of on-shell degrees of freedom for the
massless vector field (the K-R field), by saturating its propagator with
external currents $j^{\mu }$, which satisfies the following conservation
law: 
\begin{equation}
\varepsilon ^{\mu \nu \lambda }p_{\nu }j_{\lambda }=0.  \label{33.a}
\end{equation}
In 3D, a general current may be expanded with respect to a particular basis 
\begin{equation}
j^{\mu }=ap^{\mu }+b\tilde{p}^{\mu }+c\varepsilon ^{\mu },  \label{33b}
\end{equation}
where $p^{\mu }=\left( p^{0},\vec{p}\right) ,$ $\tilde{p}^{\mu }=\left(
p^{0},-\vec{p}\right) ,$ and $\varepsilon ^{\mu }=\left( 0,\vec{\varepsilon}%
\right) .$ However, (\ref{33.a}) implies that $b=c=0.$ Therefore, taking the
imaginary part of the residue (at the pole $p^{2}=0$) of the transition
amplitude, ${\cal A}=j_{\mu }^{*}\left( p\right) G_{0}^{\mu \nu }\left(
p\right) j_{\nu }\left( p\right) $, we obtain 
\begin{equation}
Im{\cal R}es\left( {\cal A}\right) =\lim_{p^{2}\rightarrow 0}\frac{\left|
a\right| ^{2}}{2}p^{2}=0.  \label{33c}
\end{equation}
From this result, we find that the massless pole does not propagate.

Returning to the Feynman rules, the torsion-fermion vertex is given by (\ref
{19}), whereas the $B_{\mu }-\psi $ vertex is as follows: 
\begin{equation}
V_{\mu }\left( B,\psi \right) =-\xi p_{\mu }.  \label{34}
\end{equation}

Once the Feynman rules for the theory have been derived, we turn out to
calculate the self-energy correction for the complete propagator.

Besides the usual self-energy corrections to the torsion and $B_\mu $%
-propagators, there is an additional mixed self-energy graph which contains
one $\phi $-field and one $B_\mu $-field as external lines. This 1-loop
graph induces an indirect $B_\mu -\phi $ coupling and, by virtue of this
mixing, we shall achieve the situation of dynamical mass generation.

Then, using the above Feynman rules, one arrives at the following results
for self-energy graphs of the fields: 
\begin{equation}
i\Pi _{\mu \nu }=i\frac{m\xi ^{2}}{\pi }p^{2}\omega _{\mu \nu },  \label{35}
\end{equation}
\begin{equation}
i\Pi _{\mu }=\frac{m\xi \alpha }{\pi }p_{\mu },  \label{36}
\end{equation}
where $i\Pi _{\mu \nu }$ and $i\Pi _{\mu }$ are the $B_{\mu }-B_{\nu }$ and
the mixed $B_{\mu }-\phi $ 1-loop self-energy diagrams, respectively (the
self-energy for the torsion field is the same as in (\ref{22})).

The contribution of eq. (\ref{36}) amounts to the radiative generation of a
finite mixing term of the form $\varepsilon ^{\mu \nu \lambda }\partial _\mu
B_{\nu \lambda }\phi $ in the 1-loop effective action. As we shall check
below, such a term does not spoil the spectrum, as no ghost state is shown
to be excited. This result suggests us that a term like the one above could
have already been introduced at tree-level, and its overall effect would
simply be a shift in the masses.

Hence, the total self-energy for the $B_{\mu }-\phi $ system is more
suitably set in a matrix form: 
\begin{equation}
i\Pi ^{\left( T\right) }=\left( 
\begin{array}{cc}
i\frac{m\xi ^{2}}{\pi }p^{2}\omega _{\mu \nu } & \frac{m\xi \alpha }{2\pi }%
p_{\mu } \\ 
-\frac{m\xi \alpha }{2\pi }p_{\nu } & -i\frac{m\alpha ^{2}}{\pi }
\end{array}
\right) .  \label{37}
\end{equation}
Analogously, the inverse propagator for this system is set in a diagonal
matrix form 
\begin{equation}
{\cal D}_{0}^{-1}=\left( 
\begin{array}{cc}
\left( 2\beta p^{2}\theta _{\mu \nu }+2p^{2}\omega _{\mu \nu }\right) & 0 \\ 
0 & \left( p^{2}-\mu ^{2}\right)
\end{array}
\right) {\rm {.}}  \label{38}
\end{equation}
The inverse of the modified propagator is given by 
\begin{equation}
{\cal D}^{-1}={\cal D}_{0}^{-1}+\Pi ^{\left( T\right) }.  \label{39}
\end{equation}
Replacing (\ref{37}) and(\ref{38}) in (\ref{39}), ${\cal D}^{-1}$ reads as 
\begin{equation}
{\cal D}^{-1}=\left( 
\begin{array}{cc}
2\beta p^{2}\theta _{\mu \nu }+2\left( 1+\frac{m\xi ^{2}}{2\pi }\right)
p^{2}\omega _{\mu \nu } & -\frac{im\xi \alpha }{2\pi }p_{\mu } \\ 
\frac{im\xi \alpha }{2\pi }p_{\nu } & p^{2}-\mu ^{2}-\frac{m\alpha ^{2}}{\pi 
}
\end{array}
\right) .  \label{40}
\end{equation}
The corrected propagators are finally obtained in momentum space by
inverting this matrix. So, the complete propagator for the $B_{\mu }$-field
reads as follows: 
\begin{equation}
iG_{\mu \nu }\left( p\right) =i\left\{ \frac{1}{2\beta p^{2}}\theta _{\mu
\nu }+\frac{\left( p^{2}-\mu ^{2}-\frac{m\alpha ^{2}}{\pi }\right) \left( 1+%
\frac{m\xi ^{2}}{2\pi }\right) ^{-1}}{2p^{2}\left( p^{2}-M^{2}\right) }%
\omega _{\mu \nu }\right\} ,  \label{41}
\end{equation}
where 
\begin{equation}
M^{2}=\mu ^{2}+\frac{m\alpha ^{2}}{\pi }+\frac{1}{2}\left( \frac{m\xi \alpha 
}{2\pi }\right) ^{2}\left( 1+\frac{m\xi ^{2}}{2\pi }\right) ^{-1}.
\label{42}
\end{equation}
This shows that the minimal coupling of torsion to matter is fundamental to
obtain a massive pole for the gauge field, since if $\alpha =0$ ($M^{2}=\mu
^{2}$), only the massless pole survives, as it can be seen from (\ref{41}).
Moreover, for the imaginary part of the residue at the massive pole, we
obtain a positive result, namely 
\begin{equation}
Im{\cal R}es\left( {\cal A}\right) =\frac{\left| a\right| ^{2}}{4}\frac{%
\left( \frac{m\xi \alpha }{2\pi }\right) ^{2}}{\left( 1+\frac{m\xi ^{2}}{%
2\pi }\right) ^{2}}>0,  \label{43}
\end{equation}
which ensures propagation of a physical degree of freedom. Therefore, the
indirect coupling between torsion and the K-R field in 3D, as mediated by
fermionic matter, results in the appearance of a massive propagating pole at
the physical longitudinal sector of $B_{\mu }$.

\section{Concluding Remarks}

The central idea of the present paper is to illustrate how torsion may play
a role in dynamically generating mass for 3D theories coupled to gravity.
Fermions minimally coupled to torsion pick out only the scalar component of
the latter, and their interaction provides the fermions with a 1-loop
non-vanishing mass correction that does not depend on the fermion tree-level
mass, but only on the coupling constant. For massless fermions, our
viewpoint is that no mass is dynamically generated for the fermions. On the
other hand, torsion itself may acquire a non-vanishing mass by means of
1-loop radiative correction.

As long as a K-R field is coupled to 3D-gravity via fermions, the
interesting result we get concerns the 1-loop generation of dynamics for the
longitudinal component of such a field: a massive and physical pole is
induced as a by-product of the coupling between torsion and the fermionic
matter that also couples to the 2-form gauge potential.

As a general conclusion from our analysis we state that {\it torsion sets
the stage for dynamical mass generation for bosonic fields in 3D} provided
non-minimal couplings are present.

\vspace{5mm}

{\Large Acknowledgments}

The authors are deeply indebted to O.M. Del Cima for helpful discussions and
critical comments on an earlier manuscript. R. Casana, L.M. de Moraes, and
G.O. Pires are also acknowledged for friendly discussions and helpful
suggestions. We thank CNPq for the invaluable financial support.

\end{document}